\documentclass[sn-mathphys,Numbered]{sn-jnl}

\usepackage{graphicx}%
\usepackage{tabularx}%
\usepackage{multirow}%
\usepackage{enumitem}
\usepackage{amsmath,amssymb,amsfonts}%
\usepackage{amsthm}%
\usepackage{mathrsfs}%
\usepackage[title]{appendix}%
\usepackage{xcolor}%
\usepackage{graphicx}
\usepackage{textcomp}%
\usepackage{manyfoot}%
\usepackage{float} 
\usepackage{booktabs}%
\usepackage{algorithm}%
\usepackage{algorithmicx}%
\usepackage{adjustbox}
\usepackage{algpseudocode}%
\usepackage{listings}%
\usepackage{caption}
\usepackage{url}%

\theoremstyle{thmstyleone}%
\theoremstyle{thmstyletwo}%

\theoremstyle{thmstylethree}%

\begin{document}

\title[Article Title]{A Social Network Approach to Analyzing Token Properties and Abnormal Events in Decentralized Exchanges}


\author*[1]{\fnm{Aryan} \sur{Soltani Mohammadi}}\email{aryan.soltani@ut.ac.ir}

\author[1]{\fnm{Moein} \sur{Karami}}\email{moein.karami@ut.ac.ir}

\author[1]{\fnm{Amir Pasha} \sur{Motamed}}\email{a.motamed@ut.ac.ir}

\author[2]{\fnm{Behnam} \sur{Bahrak}}\email{b.bahrak@teias.institute}

\affil[1]{\orgdiv{School of Electrical and Computer Engineering, College of Engineering}, \orgname{University of Tehran}, \orgaddress{\city{Tehran},  \country{Iran}}}

\affil[2]{\orgdiv{Tehran Institute for Advanced Studies}, \orgaddress{\city{Tehran}, \country{Iran}}}


\abstract{The properties of tokens within the Ethereum blockchain, such as their current prices, trade volumes, and potential future values, have been the subjects of numerous studies. Employing social networks and graphs, as powerful tools for modeling connections within groups or communities would provide valuable guidance for analyzing these properties. This study mainly focuses on creating and examining networks related to two major decentralized exchanges including Uniswap Version 2 (UniswapV2) \cite{uniswap} and SushiSwap \cite{sushiswap}. We have discovered that the distribution of nodes' degrees follows a power law that makes them scale-free networks, in addition, the centrality of tokens in exchange graphs provides valuable insights into their price and significance in cryptocurrency markets. These measures of centrality can be used to detect anomalies in cryptocurrency markets and prices. Notably, these networks exhibit remarkably similar structures, hinting at exciting research opportunities for modeling such networks. 
}

\keywords{Ethereum Blockchain, Social Networks Analysis,  Decentralized Exchanges, ERC20 Tokens}



\maketitle

\section{Introduction}\label{sec1}

Cryptocurrencies have been a fascinating innovation as they defined a new medium of currencies that function as completely decentralized, utilizing cryptography. The intervention of governments and authorities, therefore, as third parties would almost fade. This decentralized property is not limited to currencies, and there is a huge effort to find a proper decentralized alternative for any finance application. Exchange markets for cryptocurrencies, stand as one of the most significant of these applications. There are numerous centralized exchanges (CEXes) such as Coinbase \cite{coinbase} and Binance \cite{binance}, but exchanging in these markets is not anonymous, hence, decentralized exchanges (DEXes) which utilize smart contracts \cite{buterin2014next} by having such decentralized property would bring a great novelty. Smart contracts in the Ethereum blockchain play a pivotal role in designing DEXes, to be more specific a lot of famous DEXes are created with the great help of smart contracts including Uniswap and SushiSwap.

Our analyses are based on Uniswap and SushiSwap, two controversial decentralized exchanges. 
Uniswap is one of the Decentralized finance (DeFi) products as it uses smart contracts to facilitate the trades. It works within liquidity pools, and functions based on constant-product using Automated Market Makers (AMM). Each of these Uniswap liquidity pools are trading venue for a pair of ERC20 tokens \cite{vogelsteller2015eip}. The liquidity of these pools is provided by some users called liquidity providers, and some percentage of any trade that is made through this DEX will go to liquidity providers. Sushiswap as the other famous DEX is also investigated to emphasize the correctness of the properties found in Uniswap and to check the similarity of the networks generated for them. SushiSwap is a DEX that is built on the Ethereum platform and it uses the Uniswap protocol. It was at first a hard fork of Uniswap but, its code has been enhanced since, therefore some of its features vary greatly from Uniswap. 

Cryptocurrency network analysis has garnered significant research attention due to its potential to extract valuable insights from the characteristics of these currencies. Using network analysis methods, researchers can uncover hidden patterns, understand the dynamic nature of networks, and identify important nodes or groups within the digital currency system. These findings have the potential to enhance our understanding of how digital currencies work and behave, providing a solid foundation for further exploration and progress in this evolving field.
In the past two decades, network science has made substantial contributions to various scientific disciplines. The application of network analysis and graph theory, including the study of social networks, has proven to be of great help in uncovering the complexities within systems ~\cite{wasserman1994social}. 
Studying social networks and graphs has provided valuable insights into how the cryptocurrency ecosystem behaves and changes over time ~\cite{maesa2016uncovering, wu2021analysi}. 

Our primary objective is to utilize social network analysis to gain a deeper understanding of cryptocurrencies and to explore and predict trends. Unlike previous research, which predominantly examined connections between users and tokens, we aim to construct a network focused on ERC20 tokens, leveraging the intricate interactions within DEXes. This approach considering edges within tokens, to the best of our knowledge, is unique.

By creating a network that links tokens, we aim to discover fresh insights and complex patterns. This approach allows us to closely study how tokens interact in the decentralized exchange system. 
The contribution of this paper could be
summarized as follows:

\begin{enumerate}[label=\alph*.]
    \item Comparing DEX Network Structures: Compare two specific decentralized exchanges (DEXes) by analyzing their network structures to identify similarities and differences.
    \item Analyzing Token Centrality over Time: Investigate the changing centrality of tokens within DEX networks over time to understand their evolving importance.
    \item Novel Token Ranking Method: Develop and introduce an innovative method for ranking tokens based on their centrality within DEX networks, allowing for insights into token importance over defined time intervals.
    \item Identifying Network Characteristics: Identify and analyze network characteristics within DEX networks, such as the presence of power-law degree distributions and the existence of hubs, similar to social networks. 
\end{enumerate}

\section{Related work}\label{sec1}

Despite the extensive research conducted on blockchain networks of cryptocurrencies such as Bitcoin, which has greatly benefited from its long-standing deployment, the analysis of the Ethereum network has received comparatively limited attention. Existing studies on Ethereum focus primarily on analyzing transactional data, quantities, network centrality, and degree distribution~\cite{wong2022characterising}.
Chen et al. \cite{chen2020understanding} conducted a comprehensive investigation into Ethereum, utilizing graph analysis to characterize key activities such as money transfers, smart contract creation, and smart contract invocation. They introduced innovative cross-graph analysis techniques to address security concerns within the Ethereum ecosystem. Somin et al. \cite{somin2018network} focused on scrutinizing the network properties of the ERC20 network, treating trading wallets as nodes and constructing edges through buy-sell transactions. Their findings highlighted robust power-law characteristics, aligning with established expectations in network theory. 

Lin et al. \cite{lin2020modeling} adeptly modeled Ethereum transaction records as a complex network, incorporating temporal and transaction amount attributes. They devised flexible temporal walk strategies to represent this network within a random-walk-based graph framework. Meanwhile, Said et al. \cite{said2021detailed} conducted a thorough exploration of diverse facets within the Ethereum network, tracking the evolution of transactional data through a graph analysis lens. Additionally, Motamed and Bahrak \cite{motamed2019quantitative} delved into various structural attributes of five cryptocurrencies, including Ethereum, unveiling the evolving nature of transaction graphs over time and engaging in discussions on their dynamic behaviors. 

Trading cryptocurrencies due to the increase in their variety has received considerable attention. Fang et al. \cite{fang2022cryptocurrency} have written a comprehensive survey in this regard. In addition to trading them, analysis of their properties like price is being studied by researchers due to their complexity and importance. Setiawan and Bahodori \cite{setiawan2023spotting} looked at how closely the largest cryptocurrencies are tied to the regular stock market. Another area of research involves predictive modeling using network analysis. Agarwal et al. \cite{agarwal2021detecting} employed temporal graphs to identify malicious accounts in permissionless blockchains.


\section{Methodology}\label{sec3}

\subsection{Dataset}\label{subsec2}

The data used for these analyses were extracted from the Ethereum Blockchain, specifically spanning transfers between block numbers $10,060,850$ and $15,076,596$. The study focused on transfers involving liquidity pools associated with UniswapV2 and SushiSwap. These transfers represent either swap activities or the supply of liquidity by liquidity providers for the respective pools.

The approach for extracting the pools for Sushiswap is the same due to the fact that it is a hard fork of Uniswap. Thus, the focus will be on discussing the approach for Uniswap. The Uniswap Factory contract \cite{uniswap-v2-docs} was utilized for extracting pools existing in UniswapV2, which is responsible for the creation and addition of new liquidity pools to the blockchain. Two functions from this contract factory were employed to extract the pool information. The \textit{allPairsLength()} function was used to find the number of pools in the platform, and the \textit{allPairs()} function was used for identifying and recording the exact addresses of each pool on both exchanges.

\subsection{Constructing Networks} 
\label{subsubsec2}

Various networks have been created using data extracted from the Ethereum blockchain. These networks are defined as tuples of $(V, E, T, P)$, where:
\begin{itemize}
  \item $V$ represents the set of nodes, with each node corresponding to a token on either the Uniswap or SushiSwap platform.
  \item $E$ denotes the set of edges, which represent the existence of at least one transfer between tokens during specific time intervals.
  \item $T$ indicates the time interval for the transfers, which are categorized into distinct segments. For instance, $T$ could take values like $t_{i}$, where $0 \le i \le 100$. This segmentation involves splitting the block range of transfers into 100 equal segments, with $t_{0}$ referring to the entire duration of transfers.
  \item $P$ specifies whether the network is derived from Uniswap or Sushiswap.
\end{itemize}

In these networks, the presence of an edge signifies the occurrence of at least one transfer between tokens within the specified time interval, and the edges may be weighted or unweighted, depending on the network type.

\subsubsection{Weighted Network} 

Two instances of weighted networks have been created, denoted as $GW_{uni}$ and $GW_{sushi}$. In these graphs, $T$ is set to $t_{0}$, and the weight of each edge represents the number of transfers made within the respective pool.

\subsubsection{Unweighted  Network} 

To simplify the computation of metrics like density and average degree, we recognized that using a weighted network might not be accurate. Therefore, we improved our analysis by creating an unweighted network. We generated two instances of these graphs, labeled as $GU_{uni}$ and $GU_{sushi}$. 

In these graphs, represented by $T = t_{0}$, it's clear that edges in these graphs are unweighted.

\subsection{Networks Over Time}

To capture the dynamic evolution of networks over time, we have generated networks at various time intervals. From these network types, we created a total of 200 instances, with 100 dedicated to Uniswap and the other 100 for SushiSwap. These instances are represented as $GT_{uni}(t_{i})$ for Uniswap and $GT_{sushi}(t_{i})$ for SushiSwap. 

In this case, $T$ corresponds to $t_{i}$, and the edge weights are determined by the number of transfers within the pools, similar to the weighted network.

\subsection{Metrics}

The analysis of the mentioned networks was conducted by calculating various metrics. These metrics were computed using Gephi \cite{bastian2009gephi} and Python. The Networkx library \cite{hagberg2008exploring} was employed to create networks from the filtered transfers and to compute various metrics. 

Gephi was used to validate the computed values from Networkx and for visualizing the networks. Below are the definitions of the metrics utilized in this paper.

\subsubsection{Average Degree}

The average degree measures the average number of edges connected to a specific node within the network. It is calculated as follows:

\begin{equation}
Avg_{deg}(G) = \frac{2|E|}{|V|}
\end{equation}

\subsubsection{Graph Density}

The graph density metric indicates the ratio between the actual number of edges in the graph and the maximum possible number of edges it could have. It serves as a measure of the graph's sparseness, with lower values indicating a sparser graph. This metric is applied to unweighted graphs.

The density ($D$) is calculated as:

\begin{equation}
D(G) = \frac{2|E|}{{n \choose 2}}
\end{equation}

\subsubsection{Connected Components}

Connected components are subsets of nodes within the graph where there exists a path between every pair of nodes, and the addition of any node to this subset would disrupt the connectivity between pairs.

\subsubsection{Graph Diameter}

The graph diameter is a metric applicable to connected graphs. It represents the maximum number of edges within the shortest path between nodes in the graph.

\subsubsection{Degree Distribution}

The degree distribution of a graph refers to the probability distribution of node degrees across the entire network.

\subsubsection{Power Law}

A power law describes a relationship between two variables in which a simple change in one variable results in a proportional change in the other variable, raised to the power of a constant factor. This relationship is often expressed as:

\begin{equation}
f(x) \propto a \cdot (c \cdot x)^{-k}
\end{equation}

In this equation, $a$, $c$, and $k$ are constants, $x$ represents the first variable, and $f(x)$ represents the second variable.

\subsubsection{Eigenvector Centrality}

Eigenvector centrality is a fundamental network metric used to assess the influence and importance of nodes in a network. It is grounded in principles from linear algebra and relies on the idea that nodes with high eigenvector centrality are connected to other nodes of high centrality, indicating their significance within the network \cite{bonacich2001eigenvector}.

The eigenvector centrality for a given graph $G := (V, E)$ is calculated as follows:

\begin{equation}
x(v) = \frac{1}{\lambda} \sum_{u \in N(v)} x(u) \cdot w(v, u)
\end{equation}

In the context of eigenvector centrality for a node $v$, $x(v)$ represents the centrality score of that specific node. The calculation involves considering the centrality scores of neighboring nodes in the set $N(v)$, where $N(v)$ represents the neighbors of node $v$ in the graph. Additionally, the formula takes into account the edge weights that connect node $v$ to each of its neighbors $u$, denoted as $w(v, u)$.

\subsubsection{Implementation}

The code is available at \href{https://github.com/Arysoltani/Dex-Transactions-Graph}{Dex-Transactions-Graph}, and the additional results accordingly at \href{https://github.com/Arysoltani/Dex-Transactions-Graph/tree/main/Without-Weight-and-Weighted-Graph/Results}{Results}.

\section{Results}\label{sec4}

The results of our research, including network features and the correlation of significant real-world events with these networks, are discussed in this section.

\subsection{Significance of Centrality}

Node centrality within these networks reflects their importance and their centrality in the market. In this study, eigenvector centrality was chosen due to its ability to assign these values globally, meaning that the centrality of each node depends on the entire graph structure.

As shown in Table~\ref{tab:table1}, WETH (Wrapped Ether) is the most central token in both exchanges, with a significant lead over others. This result was expected, as these protocols operate on the Ethereum blockchain, which employs Ether as its primary currency. Additionally, USDT, USDC, and DAI are among the top five central tokens on both exchanges. This is attributed to the price stability of these tokens, which encourages users to exchange their assets for these tokens when seeking to minimize the risk of financial loss.

\newcolumntype{P}[1]{>{\centering\arraybackslash}p{#1}}
\newlength{\halftextwidth}
\setlength{\halftextwidth}{\dimexpr 0.4\textwidth\relax}
\newlength{\anotherhalftextwidth}
\setlength{\anotherhalftextwidth}{\dimexpr 0.7\textwidth\relax}

\begin{table}[h]
    \centering
    \caption{Top 5 tokens based on eigenvector centrality in $GW_{uni}$, $GW_{sushi}$}
    \begin{tabular}{ | P{\halftextwidth} | P{\halftextwidth} | } 
        \hline
        $GW_{uni}$ & $GW_{sushi}$\\
        \hline
        WETH : 0.991 & WETH: 0.721  \\
        \hline
        USDT : 0.088 & USDC: 0.415  \\
        \hline
        USDC : 0.085 & USDT: 0.317  \\
        \hline
        DAI : 0.032 & SUSHI: 0.232  \\
        \hline
        SHIB : 0.017 & DAI: 0.232  \\
        \hline
    \end{tabular}
    \label{tab:table1}

\end{table}

In our study, we observed that using this centrality measure for graphs generated over short time intervals provides insights into the market status during those periods. To achieve this, we employed $GT_{uni}(t_{i})$ and $GT_{sushi}(t_{i})$ for $1 \leq i \leq 100$. One of the key insights that this approach offers is the ability to detect abnormalities.

\begin{figure}[h]
    \center
    \includegraphics[width=0.5\textwidth]{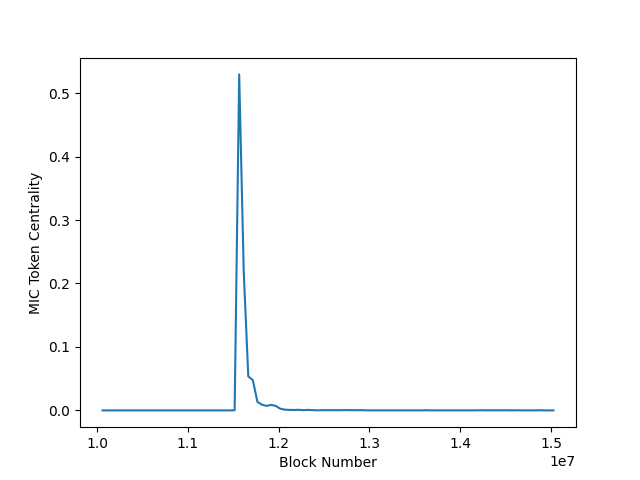}
    \caption{Centrality of MIC over time in SushiSwap}
    \label{fig:mic-centrality}
    
\end{figure}

For instance, in $GT_{sushi}(t_{i})$, the topmost tokens are usually repetitive and similar to those presented in Table~\ref{tab:table1}. However, an exception occurs in $GW_{sushi}(t_{31})$, as seen in Table~\ref{tab:table2}, where MIC and MIS tokens \cite{mithril} unusually became central. The time of these blocks is approximately January 2021, which aligns with the initial deployment of these tokens' contracts. This indicates that at the time of their introduction, the transaction volume for these tokens was significantly high, likely due to their novelty, as illustrated in Figure~\ref{fig:mic-centrality}.

\begin{table}[h]
     \caption{Top 5 tokens based on eigenvector centrality in $GW_{sushi}(t_{31})$}
    \centering
    \begin{tabular}{ | P{\anotherhalftextwidth} |} 
        \hline
        USDT: 0.682 \\
        \hline
        MIC: 0.529 \\
        \hline
        WITH : 0.386  \\
        \hline
        MIS: 0.273 \\
        \hline
        SUSHI: 0.086 \\
        \hline
    \end{tabular}
    
    \label{tab:table2}
\end{table}

Another intriguing aspect of this measure is the similarity in token centrality across different DEXs. To observe this more clearly, the centrality of these tokens has been normalized by dividing it by their average centrality. 

\begin{figure}[H]
\centering
\includegraphics[width=.45 \textwidth, height=1.7in]{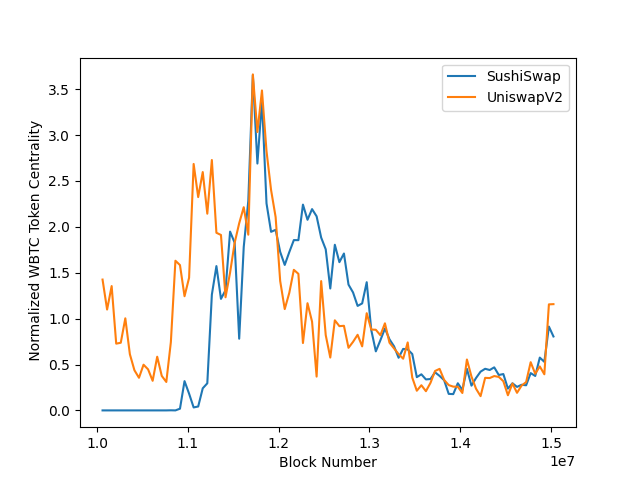}
\includegraphics[width=.45 \textwidth, height=1.7in]{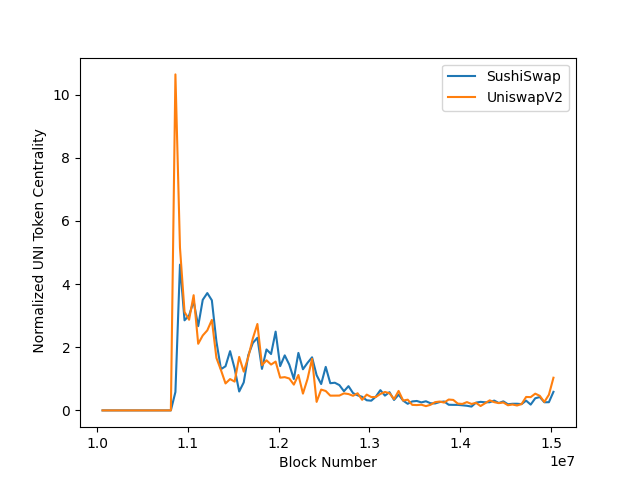}
\includegraphics[width=.45\textwidth, height=1.7in]{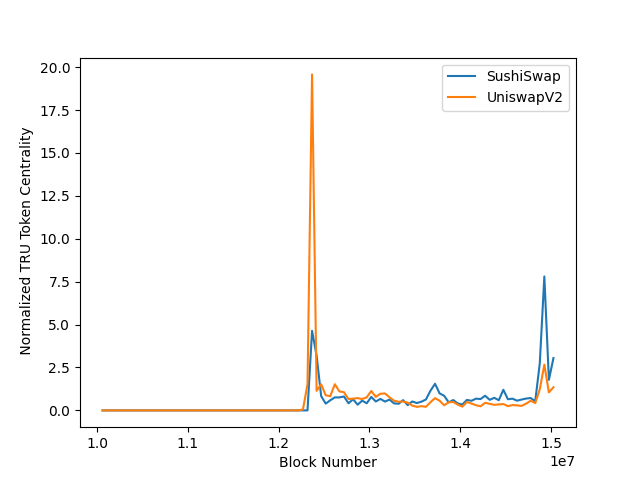}
\includegraphics[width=.45\textwidth, height=1.7in]{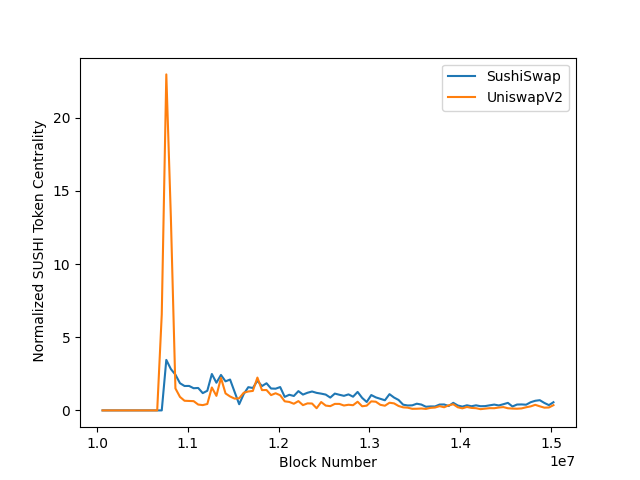}
\caption{Centrality of four tokens, including WBTC, SUSHI, UNI, and TRU, over time}
\label{fig:four_central}
\end{figure}

In Figure~\ref{fig:four_central}, we've plotted the normalized centrality of four tokens over time in both platforms. Notably, the plots for the same tokens in both DEXs closely resemble each other. This suggests the similarity of network structures in both platforms and the potential correlation with token prices.

\subsection{Networks Features}

Features of the graphs have been discussed only for networks built within $t_{0}$, which represents the time interval of all transfers used in this study. The properties were calculated based on their context and significance, which determined whether they should be calculated for weighted or unweighted graphs.

$GU_{uni}$ comprises 71,547 nodes and 76,859 edges, while $GU_{sushi}$ consists of approximately 2,400 nodes and 2,911 edges. Based on these properties:

\begin{equation}
D({GU_{uni}}) = \frac{76859}{{71547 \choose 2}} = 0.00003
\end{equation}
\begin{equation}
D({GU_{sushi}}) = = \frac{2911}{{2400 \choose 2}} = 0.001
\end{equation}

These low densities indicate the sparseness of both networks, suggesting a low likelihood of the existence of pools between tokens.

In $GU_{uni}$, $WETH$ alone is connected to 69,317 nodes, and in SushiSwap, this number is 2,000. This demonstrates that most tokens create a pool with $WETH$, and based on the number of edges, these pools comprise the majority of these edges. The average degree for $GU_{uni}$ is approximately 2.15, while for $GU_{sushi}$, it's close to 2.42.

Both $GU_{uni}$ and $GU_{sushi}$ are disconnected, consisting of a giant component \cite{erdHos1960evolution} and several smaller components. As expected, the most central nodes are part of the giant components, including $USDT$, $USDC$, $WETH$, and $DAI$. $GU_{uni}$ contains 62 components, with the giant one having a size of 71,399, while $GU_{sushi}$ has 16 components, with a giant component of size 2,369.

The diameter has been calculated for the giant component. The approximate diameter for $GU_{uni}$ is about 7, and for $GU_{sushi}$, it's about 5. Due to the fact that the number of nodes in both cases is close to the logarithm of the number of nodes in graphs, they both exhibit the "small-world property" \cite{watts1998collective}.

\subsection{Power Law Degree Distribution}

The degree distribution of both $GU_{uni}$ and $GU_{sushi}$ closely follows a power-law distribution, which is indicative of their similarity to social networks. To demonstrate this, the distributions have been plotted on a logarithmic scale. The linearity of a log-log plot of a function is commonly considered a signature of a power-law distribution. This can be confirmed by taking the logarithm of both sides of the equation:

\begin{equation}
\begin{aligned}
f(x) &\propto a \cdot (c \cdot x)^{-k} \\
\log(f(x)) &\propto \log(a) + (-k \cdot (\log(c) + \log(x))) \\
d &= \log(a) - (k \cdot \log(c)) \\
\log(f(x)) &\propto d + (-k) \cdot \log(x)
\end{aligned}
\end{equation}

\begin{figure}[H]
\centering
\includegraphics[width=.47 \textwidth, height=2.5in]{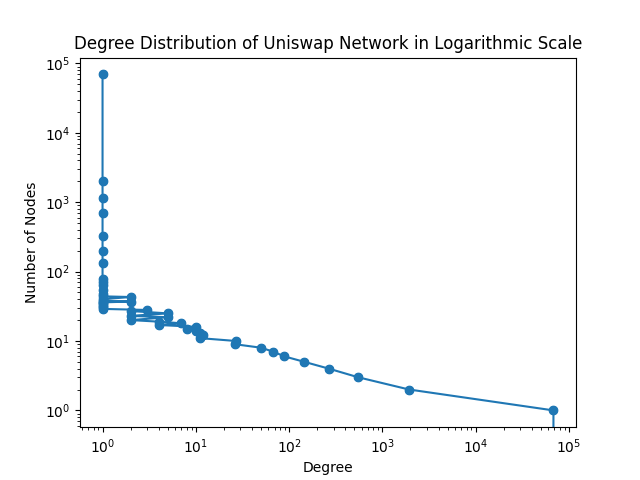}
\includegraphics[width=.47 \textwidth, height=2.5in]{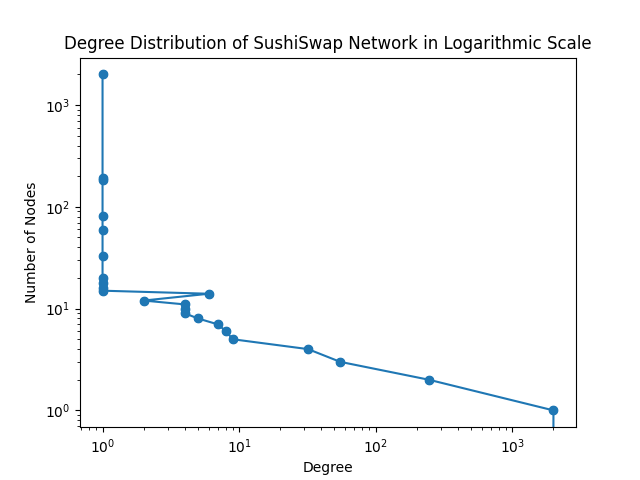}
\caption{Degree Distribution Plot of DEXes}
\label{fig:distribution}
\end{figure}

The degree distribution of both $GU_{uni}$ and $GU_{sushi}$, as depicted in Fig.~\ref{fig:distribution}, closely resembles a straight line. To further verify this observation, a hypothesis test has been designed as follows:

H(0): There is no linear relation between $x$ and $f(x)$.

H(A): These two variables have a linear relationship.

To calculate the p-value, an Ordinary Least Squares (OLS) regression model has been employed. For Uniswap, the p-value was $1.9^{-9}$, and for SushiSwap, this value was approximately $0.0005$. Based on these p-values, we conclude that these distributions exhibit a linear relationship, supporting the notion of a power-law distribution.

\subsection{Number of Nodes and Edges Evolution}

An interesting discovery is that the ratio of the number of nodes to edges in $GT_{sushi}(t_{i})$ and $GT_{uni}(t_{i})$ is the same. To verify this observation, we calculated the average and variance of the "Ratio" function defined below:

\begin{equation}
Ratio(GW_{plat}(t_{i})) = \frac{|E(GW_{plat}(t_{i}))|}{|V(GW_{plat}(t_{i}))|} 
\end{equation}

This metric represents the ratio between nodes and edges in a graph built at time $t(i)$. As expected, the variance for both Uniswap and SushiSwap is very close to $1$, with a variance of $0.006$ for Uniswap and $0.167$ for SushiSwap. Additionally, the average for Uniswap is approximately $1.2$, while for SushiSwap, it's around $1$. These findings provide evidence of the structural similarity between these networks across platforms, demonstrating that the number of nodes and edges remains consistent. It's worth noting that the relationship between new nodes and the high probability of creating edges with WETH is dependent on this fact.

\begin{figure}[H]
\centering
\includegraphics[width=.47 \textwidth, height=2.5in]{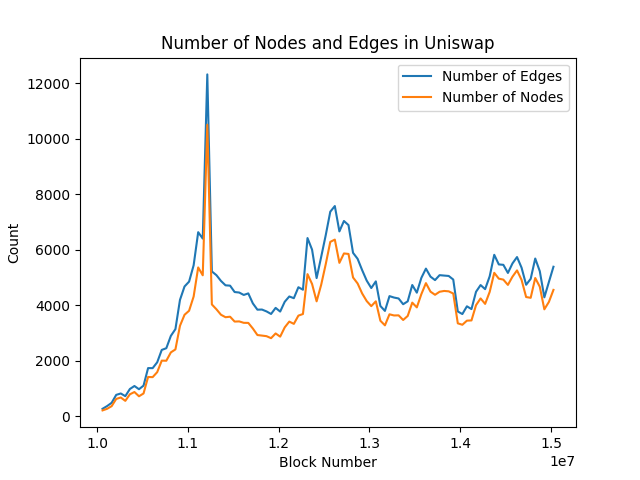}
\includegraphics[width=.47 \textwidth, height=2.5in]{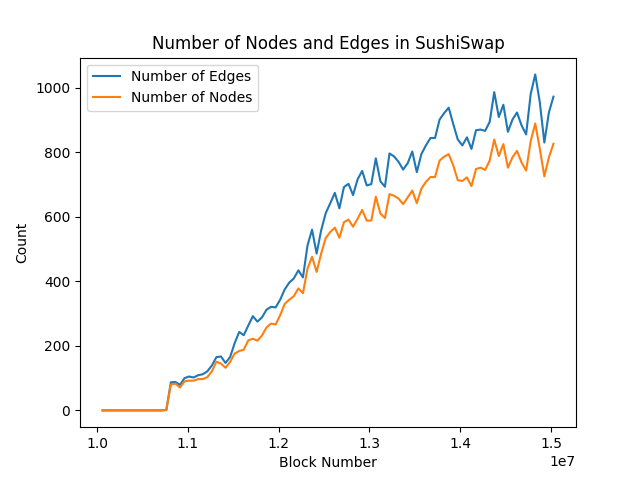}
\caption{Number of Nodes and Edges During Time}
\label{fig:Number-of-Node-and-Edges-During-Time}
\end{figure}


\subsection{Visualization of Networks}

\begin{figure}[h]
    \centering
    \includegraphics[width=0.6\textwidth]{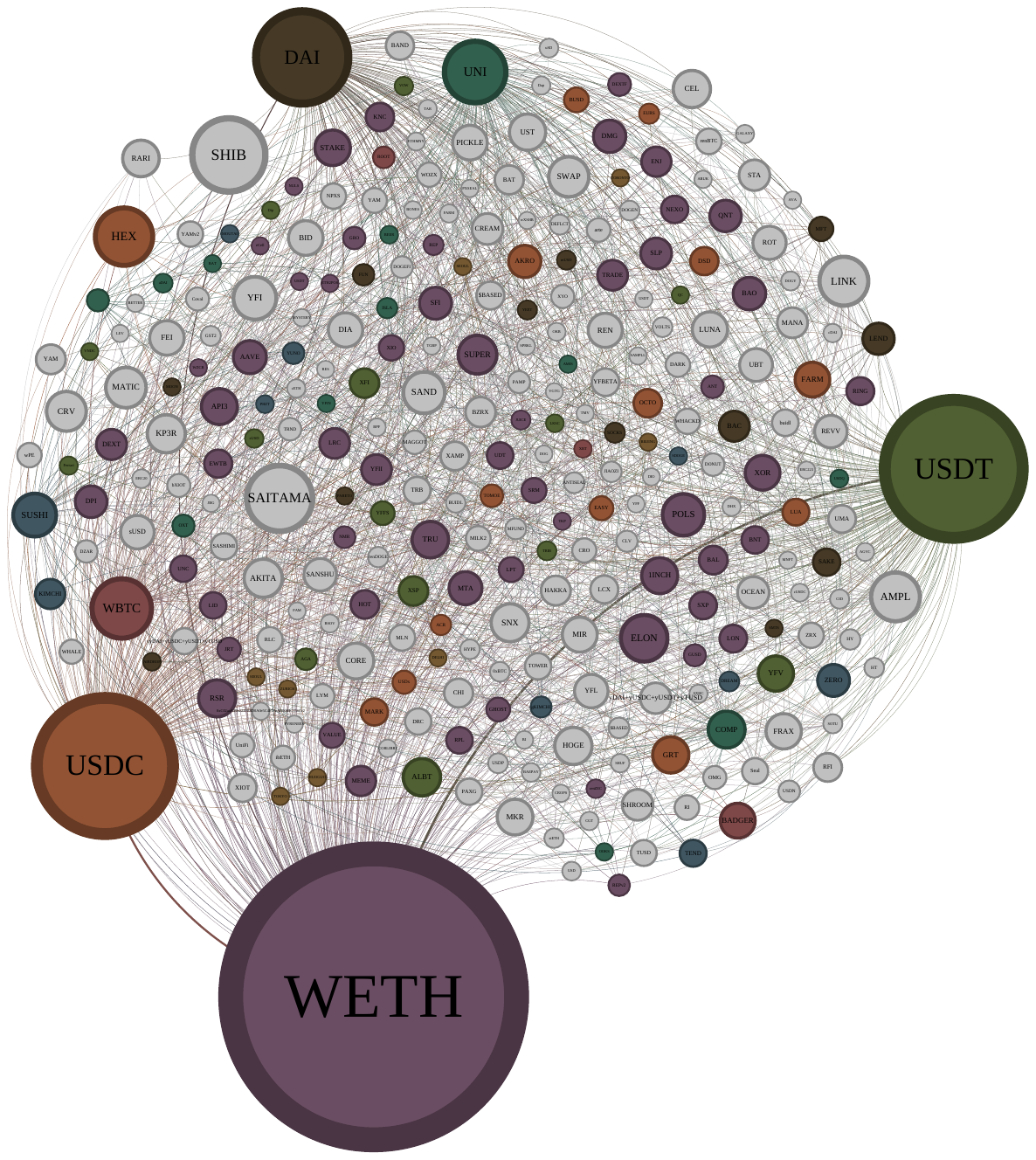}
    \caption{Visualization of Weighted Uniswap Graph}
    \label{fig:uni-vis}
\end{figure}

\begin{figure}[h]
    \centering
    \includegraphics[width=0.7\textwidth]{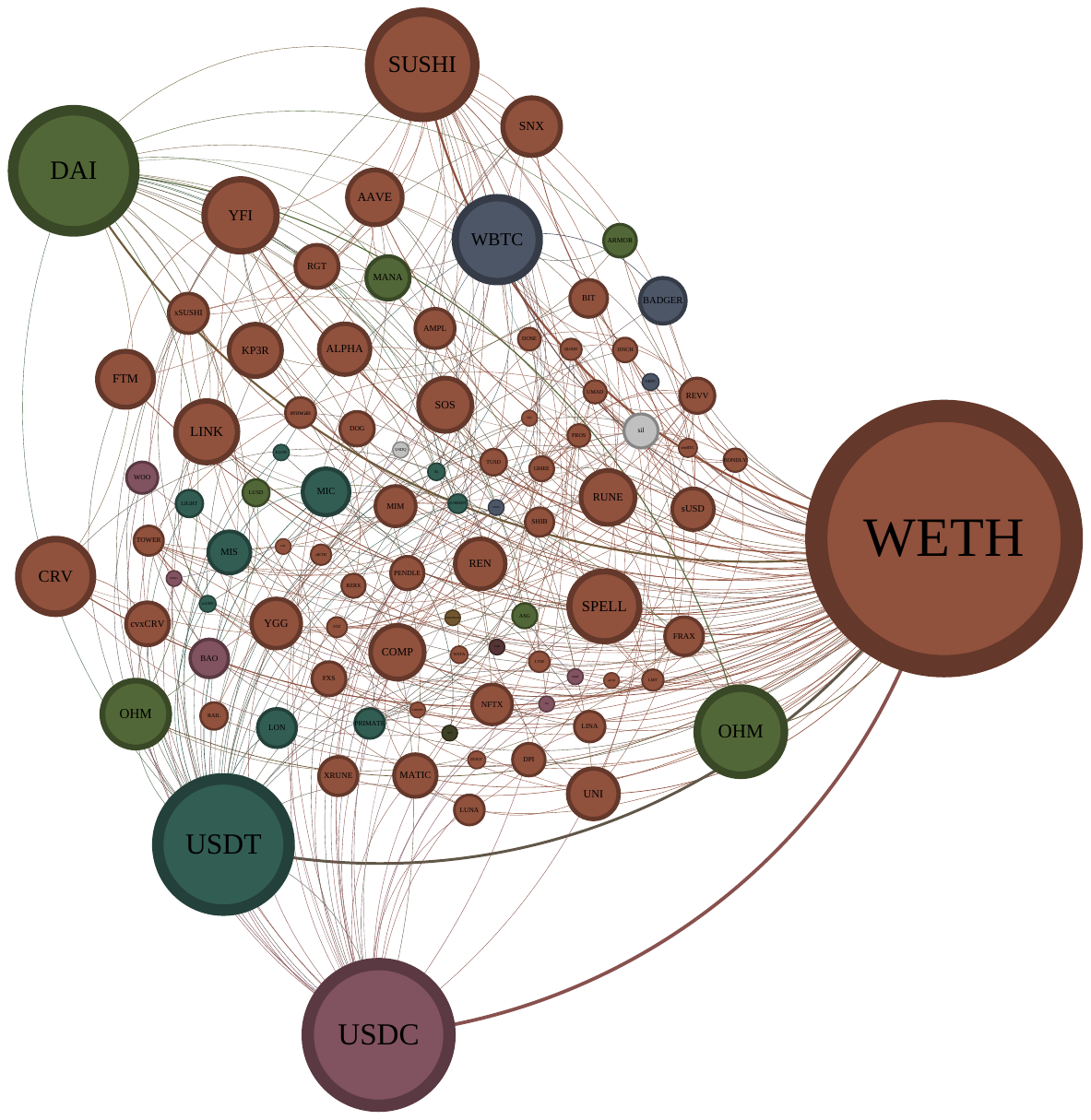}
    \caption{Visualization of Weighted SushiSwap Graph}
    \label{fig:sushi-vis}
\end{figure}

To create more manageable visualizations of the network, we utilized Gephi, a powerful network analysis and visualization tool. To streamline the graphs and focus on relevant nodes, we applied a filter that considered only nodes with a minimum degree of $5$ in Uniswap and $3$ in SushiSwap. This filter effectively excluded nodes with low connectivity and allowed us to concentrate on nodes of greater significance. Figs.~\ref{fig:uni-vis} and~\ref{fig:sushi-vis} provide visualizations of $GW_{uni}$ and $GW_{sushi}$.

In these visualizations, the size of each node is proportional to its weighted degree. Larger nodes correspond to higher degrees, while smaller nodes indicate lower degrees. Additionally, Gephi's community detection algorithm \cite{blondel2008fast} has been employed to identify and label communities within the network. The color of each node in the visualization corresponds to its assigned community. SushiSwap exhibited 33 communities, while Uniswap had 136 communities. The similarities in structures are apparent in the figures, where the larger nodes on both platforms are predominantly the same, reinforcing this observation.

\section{Future Works}\label{sec13}

Future research could encompass the examination of networks from other exchanges, whether decentralized (DEX) or centralized (CEX), to assess the consistency of network structures. The development of models that mimic these structures could prove valuable for predicting complex token behaviors and uncovering the causes of specific events. Additionally, there is potential for research focusing on fiat currency exchanges to identify similarities and differences in their functioning.

\section{Conclusion}\label{sec13}

This study has unveiled valuable insights into the relationships between token properties and the underlying graphs constructed by considering pools in two decentralized exchanges (DEXes). Notably, a significant similarity was observed between the network structures of these DEXes, both exhibiting a power-law distribution and the "small-world" property commonly found in social networks.

Furthermore, our research has shed light on the correlation between centrality measures and market behavior, emphasizing the pivotal role of centrality in explaining various events within the cryptocurrency realm. The centrality analysis has proven to be a crucial tool for understanding how tokens affect the market.

In conclusion, this study enriches our understanding of decentralized exchange ecosystems and their influence on cryptocurrency prices. The identified correlations between token properties, network structures, and market events lay the foundation for future research and strategic decision-making in the rapidly evolving realm of cryptocurrencies.

\bibliography{sn-article}

\end{document}